\documentclass[3p,times]{elsarticle}

\usepackage{ecrc}

\volume{00}

\firstpage{1}

\journalname{Nuclear Physics A}

\runauth{}

\jid{nupha}

\usepackage{amssymb}
 \usepackage{amsthm}

\usepackage[figuresright]{rotating}

\begin{document}

\begin{frontmatter}



\dochead{}

\title{Suppression of high $p_T$ hadron spectra in $p+A$ collisions}


\author[CCNU]{Rong Xu}
\author[KEK,FIAS]{Wei-Tian Deng}
\author[CCNU,LBNL]{Xin-Nian Wang}

 \address[CCNU]{Institute of Particle Physics, Central China Normal University, Wuhan 430079, China}
 \address[KEK]{Theory Center, IPNS, KEK, 1-1 Oho, Tsukuba, Ibaraki 305-0801, Japan}
 \address[FIAS]{Frankfurt Institute for Advanced Studies (FIAS) Ruth-Moufang-Strasse 1, D-60438 Frankfurt am Main, Germany}
 \address[LBNL]{Nuclear Science Division, MS 70R0319, Lawrence Berkeley National Laboratory, Berkeley, California 94720}

\begin{abstract}
Multiple hard and semi-hard parton scatterings in high-energy $p+A$ collisions involve multi-parton correlation inside the projectile in both momentum and flavor which will lead to modification of the final hadron spectra relative to that in $p+p$ collisions. Such modification of the final hadron transverse momentum spectra in $p+A$ collisions is studied within HIJING 2.1 Monte Carlo model which includes nuclear shadowing of the initial parton distributions and transverse momentum broadening. Multi-parton flavor and momentum correlation inside the projectile are incorporated through flavor and momentum conservation which are shown to modify the flavor content and momentum spectra of final partons and most importantly lead to suppression of large $p_{T}$ hadron spectra in $p+A$ collisions at both RHIC and LHC energies.
\end{abstract}


\end{frontmatter}





In heavy-ion collisions, properties of quark-gluon plasma (QGP) can be studied via jet quenching or suppression of high $p_T$ hadrons\cite{Wang:1991xy} due to parton energy loss\cite{Gyulassy:2001nm,Wang:2002ri}. However, initial multiple parton scatterings in cold nuclei can also lead to nuclear modification of the final hadron spectra at high-$p_T$. Besides the parton shadowing or nuclear modification of parton distributions inside large nuclei, multiple parton correlations inside the projectile nuclei is also an important cold nuclear effect. Energy-momentum and valence quark number conservation alone could modify the momentum and flavor dependence of final-state parton and hadron spectra at large $p_T$. Hadronization of multiple-jet systems from multiple scattering in $p+A$ collisions could also affect final hadron spectra at intermediate and large $p_T$. These cold nuclear effects need to be understood for the study of QGP properties through jet quenching.

Recently, we find that parton shadowing, initial and final-state transverse momentum broadening, final-state parton flavor composition and hadronization of multiple jets, all contribute to the nuclear modification of final hadron spectra in $p+A$ collisions\cite{Xu:2012au} within the HIJING Monte-Carlo model\cite{Wang:1991hta}. HIJING is based on a two-component model for hadron production in high-energy $p+p$, $p+A$, and $A+A$ collisions. The soft and hard components are separated by a cut-off $p_0$ in the transverse momentum exchange. Hard parton scatterings with $p_T>p_0$ are assumed to be described by the perturbative QCD (pQCD), while soft interactions are approximated by string excitations with an effective cross section $\sigma_{\mathrm{soft}}$.
Nuclear modification of initial parton distributions should be considered in $p+A$ or $A+A$ collisions. HIJING2.0\cite{Deng:2010mv} employes the factorized form of parton distributions in nuclei\cite{Li:2001xa}
\begin{equation}
 f_{a/A}(x,Q^2,b)=AR_a^A(x,Q^2,b)f_{a/A}(x,Q^2)
\end{equation}
where $R_a^A(x,Q^2,b)$ is the impact-parameter-dependent nuclear modification factor.

Besides parton shadowing, the Cronin effect\cite{Cronin:1974zm} or the enhancement of intermediate $p_T$ hadron spectra in $p+A$ collisions due to multiple parton scattering should also be considered in our calculation. Multiple scattering inside a nucleus can lead to transverse momentum ($k_T$) broadening of both initial and final state partons. This $k_T$-broadening is regarded as the reason for the Cronin effect. In HIJING 2.1, we introduce a $k_T$-kick in each binary nucleon-nucleon scattering to both the initial partons of a hard scattering as well as the final-state partons. The $k_T$-kick for each scattering follows a Gaussian distribution and an energy dependence of the width,
\begin{equation}
 \langle k^2_T\rangle=0.14\mathrm{log}(\sqrt{s}/\mathrm{GeV})-0.43\mathrm{GeV}^2.
\end{equation}
After the $k_T$-kick, longitudinal momenta of partons are reshuffled pair-wise between projectile and target partons to ensure four-momentum conservation. The introduced  $k_T$-kick will also influence slightly hadron rapidity distribution. After re-tuning the the gluon shadowing parameter $s_g$ \cite{Li:2001xa}, HIJING 2.1 can describe the rapidity distribution $dN_{ch}/d\eta$ for $d+A$ collisions at $\sqrt{s}=200$ GeV. We also extropolate our calculation to LHC energy and give the prediction of  $dN_{ch}/d\eta$ for $p+Pb$ at $\sqrt{s}=4.4$ TeV\cite{Xu:2012au}. The experiment data on the charged hadron rapidity distribution at LHC can give us further constraints on the gluon shadowing.

The hard processes in HIJING are described by lowest order pQCD. In $p+A$ collisions, the single jet inclusive cross section is proportional to nuclear parton distributions $f_{a/A}(x_2,p_T^2,b)$,
\begin{equation}
\frac{\mathrm{d}\sigma^{jet}_{pA}}{\mathrm{d}y_1\mathrm{d}^2p_T}
=K \int \mathrm{d}y_2\mathrm{d}^2bt_A(b)\sum_{a,b,c}x_1f_{a/p}(x_1,p^2_T) \times x_2f_{a/A}(x_2,p^2_T,b)\frac{\mathrm{d}\sigma_{ab\rightarrow cd}}{\mathrm{d}t},
\end{equation}
where, $x_{1,2}=p_T (e^{\pm y_1} + e^{\pm y_2})/\sqrt{s}$ are the fractional momenta of the initial partons, $y_{1,2}$ are the rapidities of the final parton jets and the higher order corrections are absorbed into the $K$ factor. The nuclear thickness function is normalized to $\int \mathrm{d}^2bt_A(b)=1$. To study parton shadowing and other cold nuclear effects, we define the nuclear modification factor for final-state parton and hadron $p_T$ spectra as,
\begin{equation}
 R_{pA}(p_T)=\frac{\mathrm{d}N_{pA}/\mathrm{d}y\mathrm{d}^2p_T}{\langle N_{bin}\rangle \mathrm{d}N_{pp}/\mathrm{d}y\mathrm{d}^2p_T},
\end{equation}
where $\langle N_{bin}\rangle$ is the average number of binary nucleon-nucleon collisions in p + A collisions.

\begin{figure}
  \centering
 \includegraphics[width=0.45\textwidth]{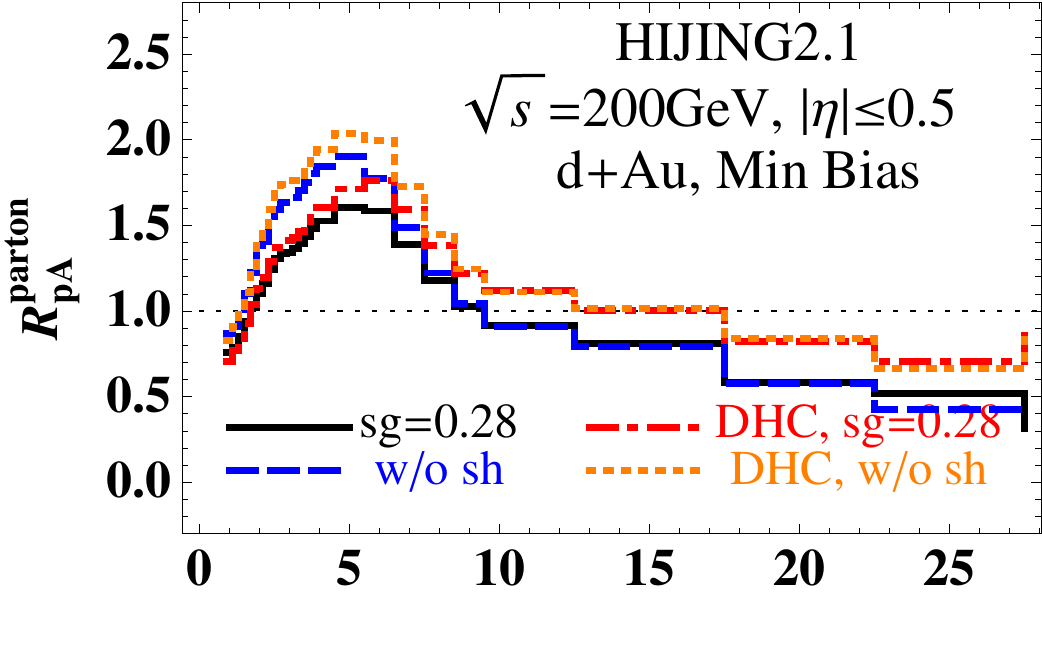}
 \includegraphics[width=0.45\textwidth]{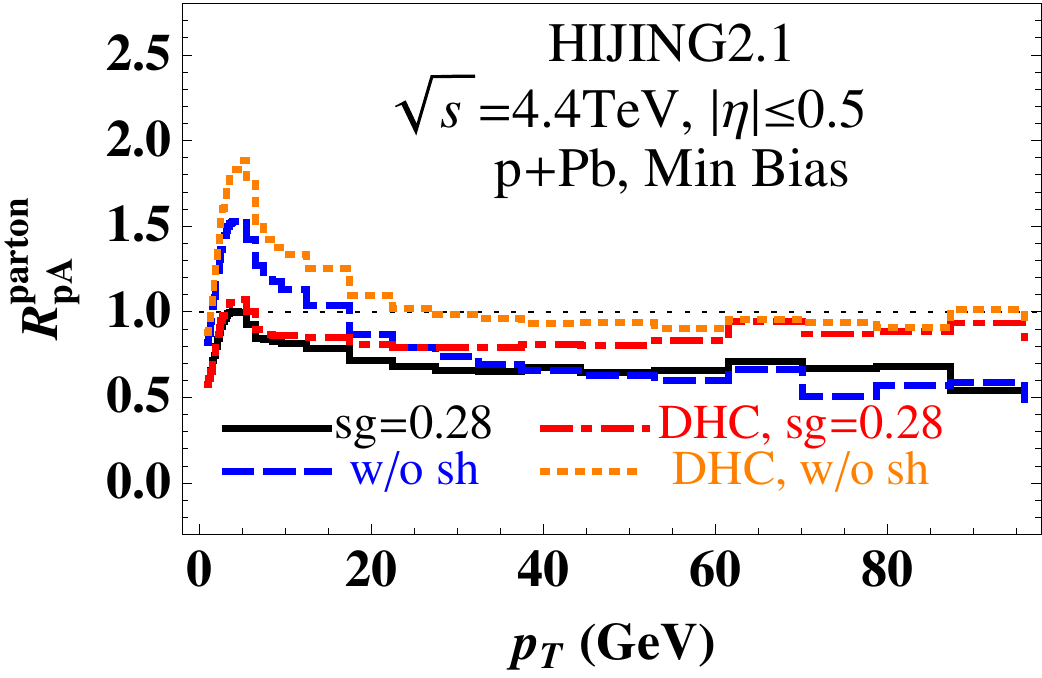}
   \caption{(color online) Nuclear modification factor for parton $p_T$ spectra in $d+A$ collisions at 200 GeV and $p+Pb$ collisions at 4.4 TeV from HIJING 2.1 model.}
  \label{fig:RpA_parton}
\end{figure}
In Fig[\ref{fig:RpA_parton}], we show the modification factor for final parton distributions $R^{parton}_{pA}$ for minimum-biased $d+Au$ and $p+Pb$ collisions at $\sqrt{s}=200$ GeV and 4.4 TeV, respectively, from the HIJING 2.1. We can see clearly the enhancement of parton spectra at intermediate $p_T$ region. In the default HIJING set, $p+A$ and $A+A$ collision are decomposed into independent nucleon-nucleon collisons. Within each nucleon-nucleon collision, hard parton scatterings are simulated first, followed by soft parton interactions. In order to maintain the energy-momentum conservation, HIJING subtracts the energy-momentum transfers in the previous hard and soft interactions from the projectile. This subtraction will restrict the energy available for subsequent binary nucleon-nucleon collisions. Since the time scale of hard scatterings is much shorter than that of soft interactions, such a coupling between hard and soft interactions within each binary collision might not be physical. In a new set of the HIJING2.1, we turn off this coupling between soft and hard parton scatterings.  We first simulate all the hard parton interactions in one $p+A$ event, then carry out all the soft interactions afterwards. As a consequence, the energy available for each hard scattering is not restricted by previous soft interactions anymore. This set is denoted as DHC (de-coherent hard scattering). In Fig[\ref{fig:RpA_parton}], we can see both the nuclear shadowing and the hard-soft coupling can suppress the $p_T$ spectra of produced partons. These features of parton spectra will be translated into the final hadron spectra after the hadronization process.

\begin{figure}
  \centering
 \includegraphics[width=0.45\textwidth]{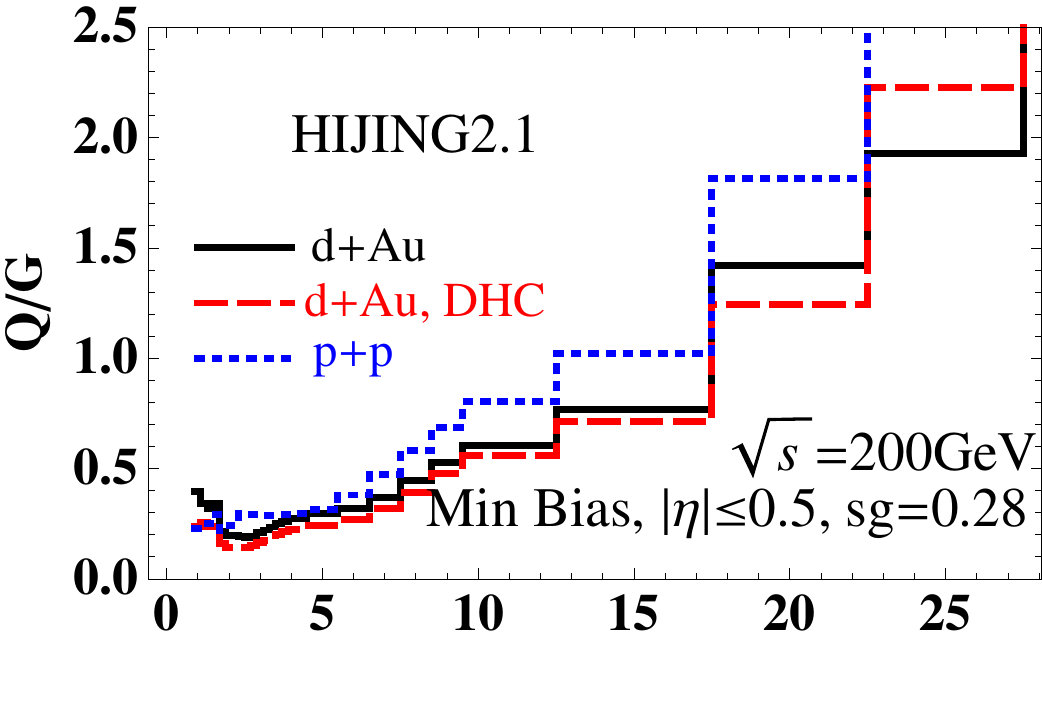}
 \includegraphics[width=0.45\textwidth]{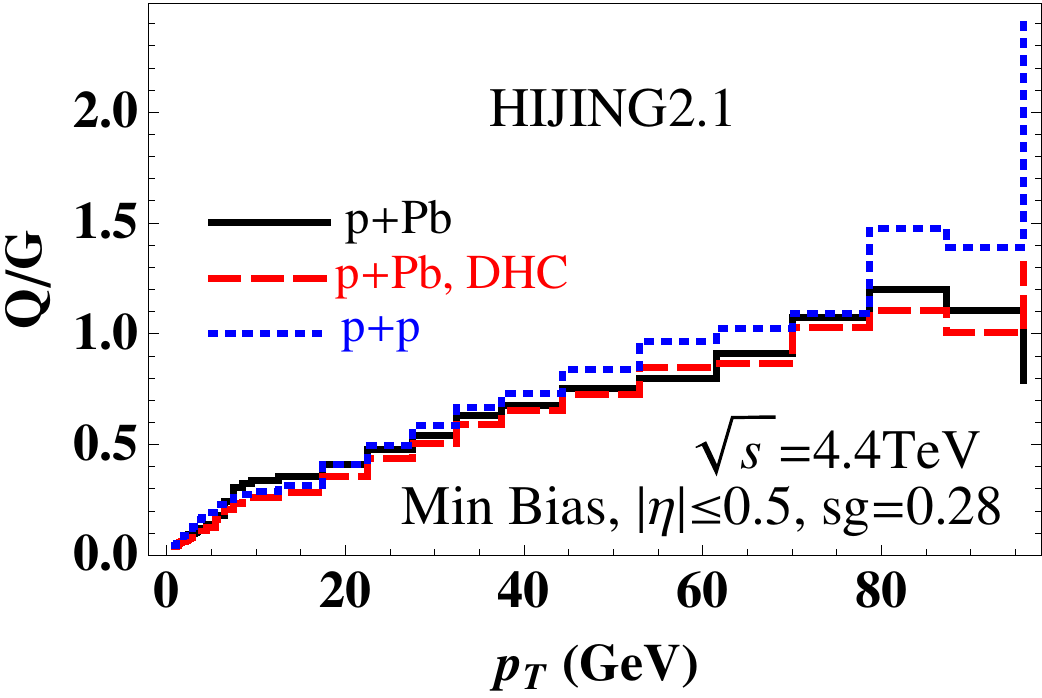}
   \caption{(color online) Quark to gluon ratio as a function of $p_T$ in $p+p$ and $p(d)+A$ collisions at 200 GeV and 4.4 TeV from HIJING 2.1 model.}
  \label{fig:R_Q/G}
\end{figure}
In $p+A$ collisions, the projectile proton will suffer multiple scatterings with the target nucleus. For each binary nucleon-nucleon collision there is a finite probability of hard parton scattering, involving independent initial partons from projectile and target nucleons. While momentum correlation of multiple partons inside the projectile can be neglected beyond the conservation of the total momentum, flavor conservation will limit the availability of valence quarks from projectile for each of these hard interactions. This valence quark number conservation will affect the flavor composition of produced partons per average binary nucleon-nucleon collision. Shown in Fig[\ref{fig:R_Q/G}] are the ratio of produced quark $p_T$ spectra over gluon, for $d+Au$ collisions at $\sqrt{s}=$ 200 GeV and $p+Pb$ collisions at $\sqrt{s}=$ 4.4 TeV, respectively, comparing with $p+p$ collisions. We can see that the flavor conservation in $p+A$ collisions can indeed suppress the fraction of quark in the produced partons relative to $p+p$ collisions, especially at high $p_T$. Because the gluon fragmentation functions are softer than quarks, the increased fraction of produced gluon jets in $p+A$ collisions can lead to suppression of final hadron spectra at high $p_T$.

Besides the momentum and flavor correlation of partons, jet hadronization or fragmentation process can also influence the final hadron spectra in $p+A$ collisions. In HIJING default set, the scattered jet partons can form jet-shower after initial and final-state radiations, and these radiated soft gluons are ordered in rapidity. While the $q-\bar{q}$ jets shower can form independent string, the gluon jet shower are always connected to the valence quark and di-quark of the corresponding projectile and target nucleons as kinks to form string systems. Within Lund string fragmentation model\cite{Andersson:1983ia}, all strings are fragmented into final hadron. In $p+A$ collisions, the projectile can undergo multiple scatterings, its string system can have more gluon attached comparing to that in $p+p$ collisions. Hadrons fragmented from such string system are softer than hadrons from independent fragmentation of individual gluons. To illustrate this effect, we also compare the results of independent fragmentation for both $p+p$ and $p+A$ collisions with HIJING 2.1.

\begin{figure}
  \centering
 \includegraphics[width=0.45\textwidth]{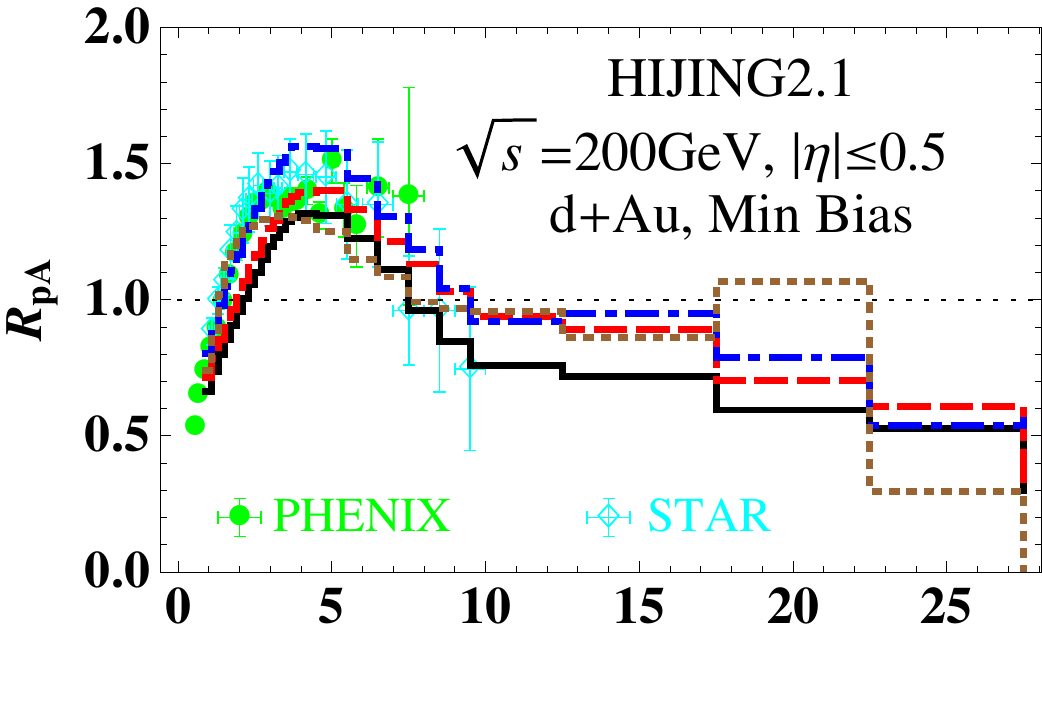}
 \includegraphics[width=0.45\textwidth]{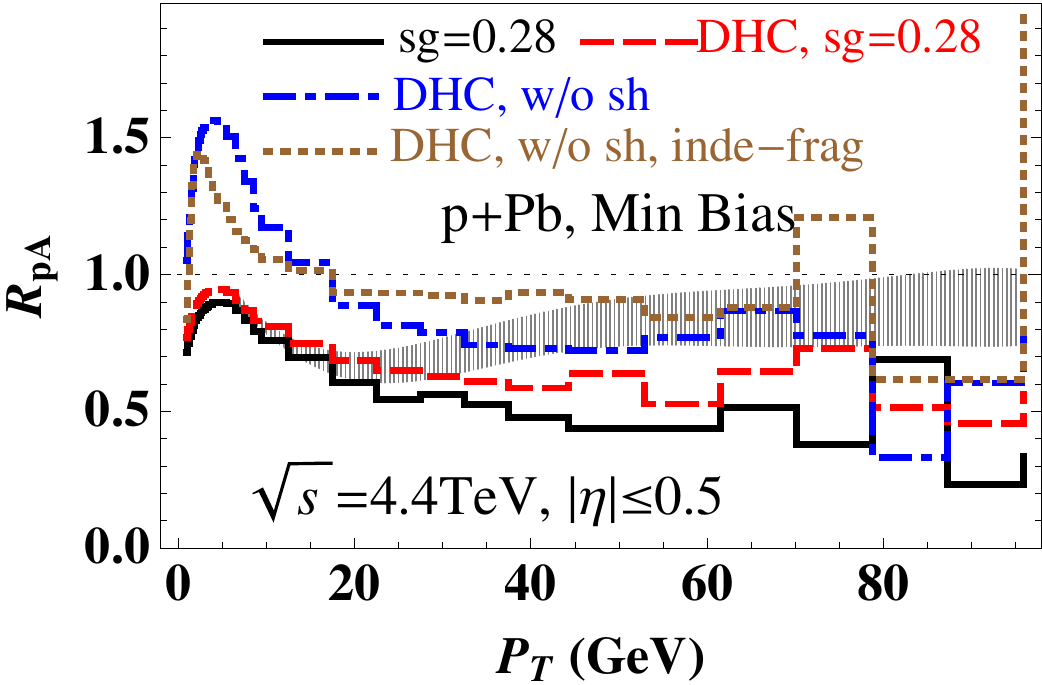}
   \caption{(color online) Nuclear modification factor for charged hadrons in $p+A$ collisions from HIJING 2.1 model, comparing with the PHENIX and STAR experiment data. See detail in text.}
  \label{fig:RpA_hadron}
\end{figure}
In Fig[\ref{fig:RpA_hadron}], we show the nuclear modification factors for charged hadrons in $d + Au$ at $\sqrt{s} = 200$ GeV and $p + Pb$ collisions at $\sqrt{s} = 4.4$ TeV, respectively, as compared to the existing data from the PHENIX\cite{Adler:2003ii} and STAR\cite{Adams:2003im} experiment at RHIC. If there is no nuclear effect, $R_{pA}\approx1$ at intermediate and large $p_T$ region. The enhancement of hadron spectra at intermediate $p_T$ is due to the $k_T$ broadening through multiple scatterings. As one can see from Fig[\ref{fig:RpA_hadron}], the parton shadowing in nuclei, soft-hard coupling and enhanced gluonic jets due to valence quark conservation all lead to suppression of charged hadron spectra. Furthermore, the fragmentation of string systems with multiple gluons from multiple hard parton scatterings can further suppress the hadron spectra in the high $p_T$ region.

Since parton shadowing will disappear at large $p_T$\cite{Eskola:2009uj} due to QCD evolution and hard scatterings are de-coherent from soft interactions, the nuclear modification factor at LHC will likely follow the default set and DHC results at low $p_T$ and approach DHC without shadowing at large $p_T$ with possible further modifications due to hadronization of multiple
jets. This possibl scenario is shown as the shaded band in Fig[\ref{fig:RpA_hadron}] for the nuclear modification $R_{pA}$ in $p+Pb$ collisions at the LHC energy.

In summary, we have studied the nuclear modification of hadron spectra in $d+Au$ and $p+Pb$ collisions at the RHIC $\sqrt{s}_{NN}=200$ GeV and LHC energy $\sqrt{s}_{NN}=4.4$ TeV within the HIJING2.1 Monte Carlo model. The nuclear modification of the $p_{T}$ spectra is found to be subjected to correlated hadronization of multiple jets and flavor conservation in multiple parton scattering in the HIJING model. Experimental test of these effects in the proposed $p+Pb$ collisions at the LHC is crucial to disentangle these cold nuclear effects from that caused by jet quenching in the hot quark-gluon plasma.

\section*{Acknowledgement}
This work was supported in part by the NSFC under the project No. 10825523,
by self-determined research funds of CCNU from the colleges’ basic research and operation of MOE,
Helmholtz International Center for FAIR within the framework of the LOEWE program launched by the State of Hesse,
US Depart of Energy under Contract No. DE-AC02-05CH11231 and within the framework of the JET Collaboration,
and Grant-in Aid for Scientific Research (No. 22340064) from the Ministry of Education, Culture, Sports, Science and Technology (MEXT) of Japan.


\bibliographystyle{elsarticle-num}






\end{document}